\documentclass[%
 reprint,
 amsmath,amssymb,
 aps,
prb,
]{revtex4-2}

\usepackage{appendix}
\usepackage{lipsum}
\usepackage{xcolor}
\usepackage{physics}
\usepackage{bbm}
\usepackage{graphicx}
\usepackage{dcolumn}
\usepackage{bm}

\begin{document}

\preprint{APS/123-QED}

\title{Optical valley separation in two-dimensional semimetals with tilted Dirac cones}

\author{Andrew Wild}
\email{A.Wild@exeter.ac.uk}
\author{Eros Mariani}
\email{E.Mariani@exeter.ac.uk}
\author{M. E. Portnoi}
\email{M.E.Portnoi@exeter.ac.uk}
\affiliation{%
 Physics and Astronomy, University of Exeter, Stocker Road, Exeter EX4 4QL, United Kingdom
}%

\date{\today}

\begin{abstract}
 Two-dimensional semimetals with tilted Dirac cones in the electronic band structure are shown to exhibit spatial separation of carriers belonging to different valleys under illumination. In stark contrast to gapped Dirac materials this optovalleytronic phenomenon occurs in systems with intact inversion and time-reversal symmetry that host massless Dirac cones in the band structure, thereby retaining the exceptional graphene-like transport properties. As a result we demonstrate that optical valley separation is possible at arbitrarily low photon frequencies including the deep infrared and terahertz regimes with full gate tunability via Pauli blocking. As a specific example of our theory, we demonstrate tunable valley separation in the proposed two-dimensional tilted Dirac cone semimetal 8-$Pmmn$ borophene for incident infrared photons at room temperature.
\end{abstract}
                       
\maketitle

\section{Introduction}

Electrons in Dirac materials behave as massless fermions, existing in one of two inequivalent Dirac cones (known as valleys) with a low energy linear electronic dispersion. As the valleys in Dirac materials are widely separated in momentum space, carriers rarely scatter between them in the absence of atomic-scale disorder. This makes Dirac materials ideal candidates for valleytronic applications where the valley index encodes quantum information. For the realization of valleytronic devices it is vital to achieve independent control over carriers with different valley indices. The best known mechanism of optovalleytronics is in materials with broken inversion and preserved time-reversal symmetry that host gapped Dirac cones in their band structure. In such systems individual valleys can be addressed with different circularly polarized photons and under an external, in-plane electric field carriers from different valleys are steered in opposite directions yielding a finite photocurrent\,\cite{PhysRevLett.99.236809,PhysRevB.77.235406,Mak2014,doi:10.1126/science.1254966}.

The aforementioned mechanism does not offer optovalleytronic applications for gapless two-dimensional (2D) Dirac fermions. Protecting the gapless nature of Dirac particles preserves their superior transport properties in the form of high mobility due to the suppression of back-scattering associated with Klein tunneling. These merits come at a cost – it becomes difficult to control the propagation of charge carriers. One solution to this problem utilizes the optical momentum alignment phenomenon in which photocarriers in Dirac materials such as graphene excited by linearly-polarized light propagate perpendicular to the polarization plane\,\cite{HartmannBook,*HartmannThesis}. Momentum alignment could be exploited for valleytronic applications in materials that exhibit a certain degree of valley anisotropy in the band structure. An example of such an anisotropy is the trigonal warping of the electronic dispersion of graphene which becomes noticeable from about 1eV above the apex of the Dirac cone\,\cite{HartmannBook,*HartmannThesis,Saroka2022}. However, this mechanism is limited to high excitation frequency preventing any control of valley separation by means of a gate voltage - the main asset of 2D materials for optoelectronic applications.

\begin{figure}[t]
\includegraphics[width=0.48\textwidth]{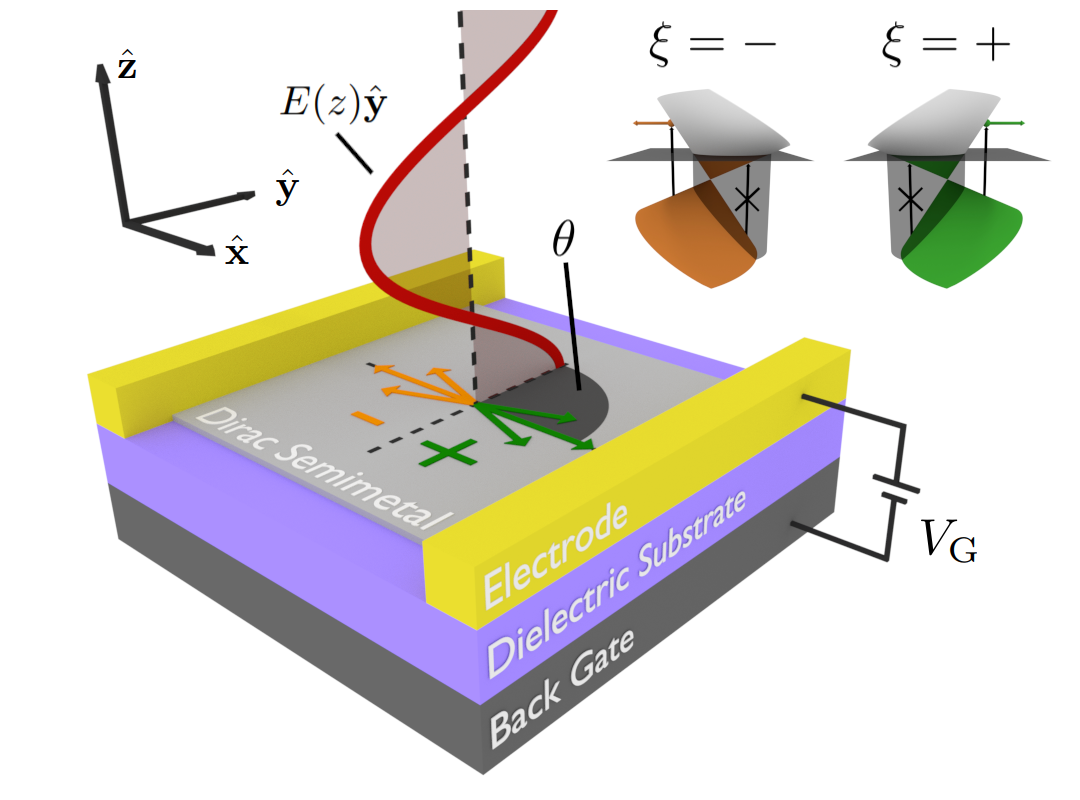}
\caption{\label{fig:Schem} Schematic of the suggested experimental setup for optically generating valley carrier separation in 2D tilted Dirac cone materials. A back-gate configuration with gate voltage $V_\text{G}$ can be used to the change the Fermi level $E_\text{F}$. Linearly polarized photons are described by an electric field which propagates along the $\hat{\textbf{z}}$ direction and is polarized at angle $\theta$ to the crystallographic $\hat{\textbf{x}}$ axis. The inset shows the band structure of two tilted Dirac cones with valley index $\xi = \pm$ (sketched in green and orange). The incident photons induce interband transitions - in the shaded regions optical transitions are Pauli blocked. The resulting group velocity of photoexcited carriers depends on their valley index.}
\end{figure}

In this work we propose a tunable mechanism of optical valley separation in high-mobility 2D semimetals over a broad range of excitation frequencies including the elusive terahertz regime. This opportunity is offered by materials hosting tilted Dirac cones in the electronic band structure where the two valleys are skewed in opposite directions (see inset of Fig.\,\ref{fig:Schem}). Combining this intrinsic valley anisotropy with optical momentum alignment and Pauli blocking effects it becomes possible to spatially separate photoexcited carriers with different valley index away from the light spot (see Fig.\,\ref{fig:Schem}). The degree of valley polarization can be controlled via Pauli blocking which in 2D semimetals is readily tuned with a back gate. The spatial separation of valley carriers results in unequal valley populations at opposite sides of the light spot. This effect can be detected by measuring the degree of circular polarization of the edge luminescence in a nearby gapped material\,\cite{PhysRevLett.99.236809,PhysRevB.77.235406,PhysRevB.103.165415}, which ideally could be the same material with locally broken inversion symmetry.

Tilted Dirac cones appear in three varieties: sub-critically tilted (type-I) with closed elliptical isoenergy contours, critically tilted (type-III) with open parabolic isoenergy contours and super-critically tilted (type-II) with open hyperbolic isoenergy contours. Two-dimensional materials hosting tilted Dirac cones are an ever growing family with candidate materials including 8-$Pmmn$ borophene\,\cite{PhysRevLett.112.085502,PhysRevB.93.241405,PhysRevB.94.165403}, an organic salt $\alpha$-(BEDT-TTF)$_2$I$_3$\,\cite{doi:10.1143/JPSJ.75.054705} and many more\,\cite{PhysRevB.78.045415,PhysRevLett.105.037203,doi:10.1126/science.1256815,PhysRevX.6.041069,PhysRevB.94.195423,Ma2016,https://doi.org/10.1002/pssr.201800081,PhysRevB.98.121102,PhysRevB.100.235401,PhysRevB.100.205102,PhysRevB.102.041109}. As a case study of our work we demonstrate tunable valley separation in 8-$Pmmn$ borophene upon illumination of infrared photons at room temperature. We further demonstrate that type-II Dirac cone materials always possess perfect optical valley separation due to their super-critically tilted band dispersion. As an extension to our theory we show that type-III Dirac cones will display enhanced momentum alignment and emission of highly polarized terahertz photons via hot luminescence aided by the inclusion of carrier scattering.

\section{Model}
We consider a 2D Dirac semimetal with tilted Dirac cones in the electronic band structure described by the Bloch Hamiltonian 
\begin{equation}
\label{eq:Hamiltonian}
    H^\xi(\textbf{q}) = \hbar v_\text{F} \big( \xi \gamma \eta q_x \mathbbm{1} + \xi \eta q_x \sigma_x + q_y \sigma_y \big),
\end{equation}
where $\sigma_x$ and $\sigma_y$ are Pauli matrices, $\mathbbm{1}$ is the $2 \times 2$ identity matrix and $v_\text{F}$ is the Fermi velocity along $q_y$ where $\textbf{q} = (q_x, q_y)$ is the wavevector measured from the Dirac point in the Brillouin zone corresponding to the inequivalent valleys $\xi = \pm$. The Dirac Hamiltonian has a tilt parameter $\gamma$ which defines sub-critically tilted ($\left|\gamma\right| < 1$, type-I), critically tilted ($\left| \gamma \right| = 1$, type-III) and super-critically tilted ($\left| \gamma \right| > 1$, type-II) Dirac cones. The anisotropy parameter $\eta > 0$ scales the Dirac cone along the tilt axis. The valley-dependent eigenenergies and eigenvectors of the Hamiltonian are defined as
\begin{equation}
\label{eq:eigenvalues}
    E^\xi_\pm(\textbf{q}) = \hbar v_\text{F} q \Big[ \xi \gamma \eta \cos(\varphi_\textbf{q}) \pm \sqrt{\eta^2 \cos^2(\varphi_\textbf{q}) + \sin^2(\varphi_\textbf{q})} \Big],
\end{equation}
and 
\begin{equation}
    \ket*{\Psi^\xi_\pm(\textbf{q})} = \frac{1}{\sqrt{2}} \begin{bmatrix}
    \pm \frac{\xi \eta \cos(\varphi_\textbf{q}) - i \sin(\varphi_\textbf{q})}{\sqrt{\eta^2 \cos^2(\varphi_\textbf{q}) + \sin^2(\varphi_\textbf{q})}} \\ 1
    \end{bmatrix}
\end{equation}
respectively, for the conduction ($+$) and valence ($-$) bands. Here we have defined the wavevector in polar coordinates as $q_x = q\cos(\varphi_\textbf{q})$ and $q_y = q \sin(\varphi_\textbf{q})$ with $q$ the radial wavevector and $\varphi_\textbf{q}$ the wavevector angle. The semimetal has a Fermi level $E_\text{F}$ that can be tuned by means of a metallic back gate as shown in Fig.\,\ref{fig:Schem}. The sample is incident upon by linearly polarized photons with polarization $\hat{\textbf{e}}_\theta = \cos(\theta)\hat{\textbf{x}} +  \sin(\theta)\hat{\textbf{y}}$ and energy $h \nu$. We treat the corresponding electric field as a time-dependent perturbation to the otherwise time-independent system using Fermi's golden rule inducing vertical, interband transitions. In this work we do not consider intraband absorption as it requires knowledge of material-dependent scattering mechanisms and in the case of type-II Dirac cone materials, a detailed understanding of the Fermi surface beyond the Dirac cone approximation. We also note that our mechanism works for photons incident normally on the sample and does not rely on in-plane momentum transfer to electrons via phenomena such as photon-drag\,\cite{GLAZOV2014101}.

There are three factors that govern the optical absorption of photons. $i)$ Initial and final states with wavevector $\textbf{q}$ must be separated by an energy of $\Delta E(\textbf{q}) = E^\xi_+(\textbf{q}) - E^\xi_-(\textbf{q}) = h \nu$. For a fixed frequency $\nu$ this condition gives a set of wavevectors available for the transition given by
\begin{equation}
\label{eq:DeltaE}
    \Delta E(\textbf{q}) = 2 \hbar v_\text{F} q \sqrt{\eta^2 \cos^2(\varphi_\textbf{q}) + \sin^2(\varphi_\textbf{q})}.
\end{equation}
It can be seen that the states contributing to absorption fall on the perimeter of an ellipse in wavevector space with semi-major and semi-minor axes ($\pi \nu / v_\text{F}$ and $\pi \nu / \eta v_\text{F}$) proportional to the frequency of the incident photon. For the case of the anisotropy parameter equaling unity ($\eta = 1$), this ellipse becomes a circle with radius $\pi \nu/v_\text{F}$. The geometry of this ellipse is independent of both the valley index ($\xi$) and tilt parameter ($\gamma$). $ii)$ The transition rate describes the likelihood of an absorption event occurring at a given wavevector. For linearly polarized photons the transition rate is proportional to the absolute value squared of the expectation value of the velocity operator projected along the axis of polarization [$v_{\text{cv}}(\textbf{q}) = \bra*{\Psi^\xi_\pm(\textbf{q})} \hat{\textbf{e}}_\theta \cdot \textbf{v} \ket*{\Psi^\xi_\mp(\textbf{q})}$] between the initial and final states\cite{Anselm,HartmannBook}. The velocity operator within the gradient approximation $\textbf{v} = (1/\hbar) \bm{\nabla}_\textbf{q} H^\xi(\textbf{q})$, leads to the squared velocity matrix element
\begin{equation}
    \mid \! v_{\text{cv}}(\varphi_\textbf{q}) \! \mid^2 \: = \frac{\eta^2 v_\text{F}^2}{\eta^2 \cos^2(\varphi_\textbf{q}) + \sin^2(\varphi_\textbf{q})}\sin^2(\varphi_\textbf{q} - \theta).
\end{equation}\begin{figure}
    \centering
    \includegraphics[width=0.48\textwidth]{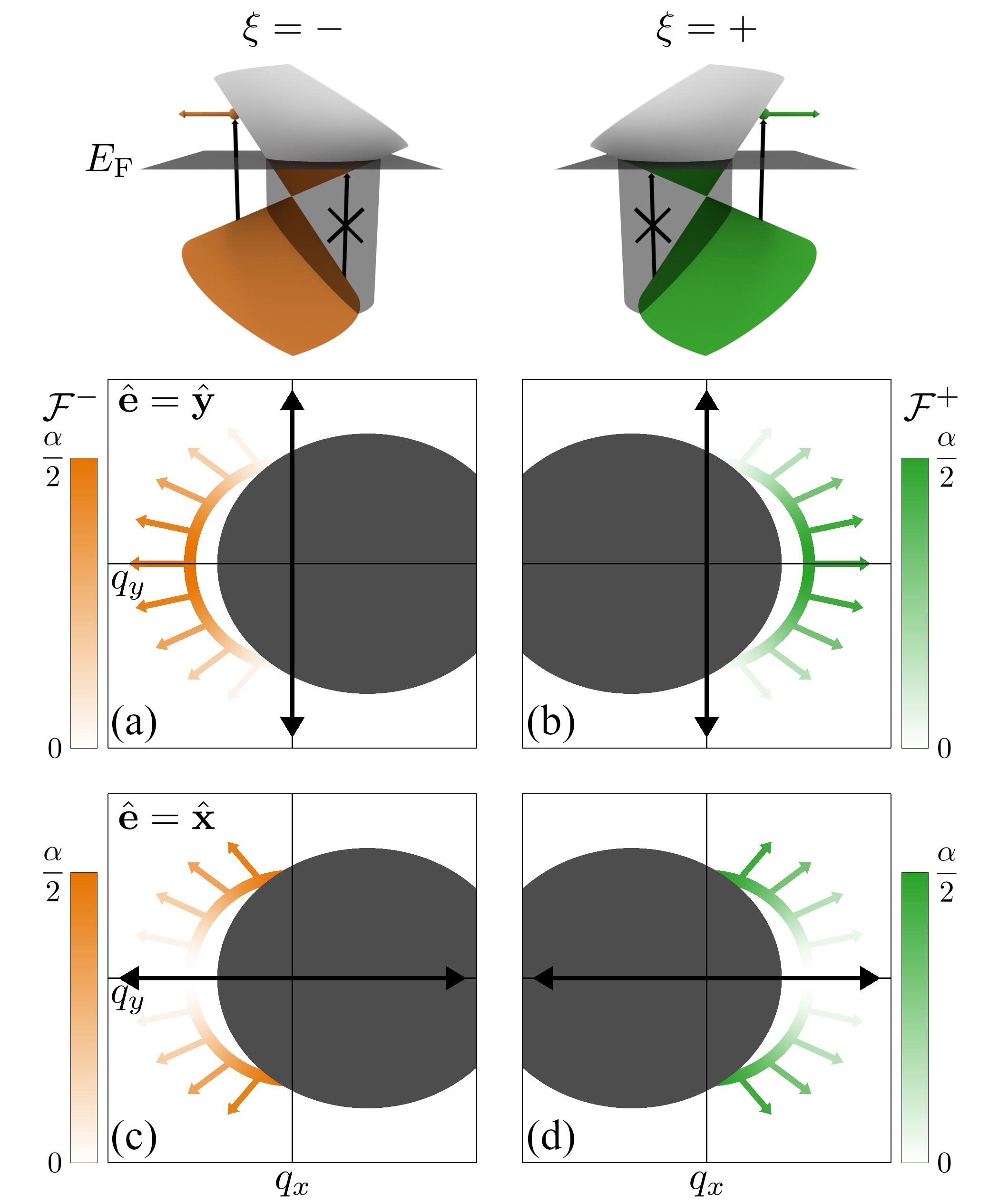}
    \caption{Distribution of photoexcited carriers ($\mathcal{F}^\xi$) in a type-I Dirac cone material with two valleys $\xi = +$ and $-$ sketched in green and orange respectively. In this plot the Dirac cone tilt is $\gamma = 0.5$, anisotropy is $\eta = 1$ and $\alpha \approx 1/137$ is the fine structure constant. The Fermi level ($E_\text{F}$) sits above the Dirac point creating regions of Pauli blocked transitions (dark gray). The group velocities of photoexcited carriers have been projected in to wavevector space and sketched as arrows. Due to Pauli blocking and the conical band structure, photoexcited carriers in the two valleys have differing group velocities. Optimal spatial separation of valley carriers is achieved when photon polarization is aligned with the crystallographic $\hat{\textbf{y}}$ axis [panels (a) and (b)] compared to the $\hat{\textbf{x}}$ axis [panels (c) and (d)].}
    \label{fig:PolarisationChange}
\end{figure}The formula above demonstrates momentum alignment in Dirac materials in which, upon absorption of photons polarized along $\theta$, carriers are generated with wavevector angle $\varphi_\textbf{q}$ predominantly perpendicular to the polarization vector. The velocity matrix element is equivalent to that of non-tilted cones and hence is independent of both valley ($\xi$) and tilt parameter ($\gamma$). $iii)$ For an absorption event to occur we must ensure that the initial state is occupied and the final state is empty to avoid Pauli blocking. We define these conditions with the Fermi-Dirac distributions $f_e(E) = 1 - f_h(E) = \big\{ 1 + \exp \big[ (E - \mu)/k_\text{B} T \big] \big\}^{-1}$ for electrons ($e$) and holes ($h$) with Boltzmann constant $k_\text{B}$, chemical potential $\mu$ and temperature $T$. Crucially, the regions of Pauli blocking are valley ($\xi$) and tilt parameter ($\gamma$) dependent leading to valley-dependent distributions of photoexcited carriers for certain values of the Fermi energy. Combining all of these factors we can write the angular distribution of excited carriers\,\cite{HartmannBook,*HartmannThesis} at the instant of photocreation as
\begin{equation}
\label{eq:Carrier}
\begin{split}
    &\mathcal{F}^\xi(\varphi_\textbf{q}) = \alpha \frac{g_\text{s} \hbar}{2 \pi \nu} \mid \! v_\text{cv}(\varphi_\textbf{q}) \! \mid^2 \times \\ &\int_{0}^\infty \!\!\!\!\!\! \delta \big[ \Delta E(\textbf{q}) - h \nu \big] f_e \big[E^\xi_-(\textbf{q})\big] f_h \big[E^\xi_+(\textbf{q})\big] q \text{d} q,
\end{split}
\end{equation}
where $g_\text{s} = 2$ accounts for the spin degeneracy, $\delta[...]$ is the Dirac delta function and $\alpha = e^2/\hbar c \approx 1/137$ is the fine-structure constant in CGS units. We note that the normalization factor is chosen such that integrating over the wavevector angle yields the ratio of absorbed photons. Accordingly, the absorption is defined as $\mathcal{A} = g_\text{v} \int_0^{2\pi} \mathcal{F}^\xi(\varphi_\textbf{q}) \text{d} \varphi_\textbf{q}$, where $g_\text{v} = 2$ is the valley degeneracy\,\cite{Wild2022}. Note that for the case of graphene ($\gamma = 0$ and $\eta = 1$) this expression simplifies to the well-known universal sheet absorption of $\mathcal{A} = \pi \alpha \approx 2.3\%$. For the full analytic expression of the distribution of photoexcited carriers $\mathcal{F}^\xi(\varphi_\textbf{q})$ see Appendix \,\ref{sec:FFormula}.

\section{Results}
\begin{figure*}[t]
    \centering
    \includegraphics[width=\textwidth]{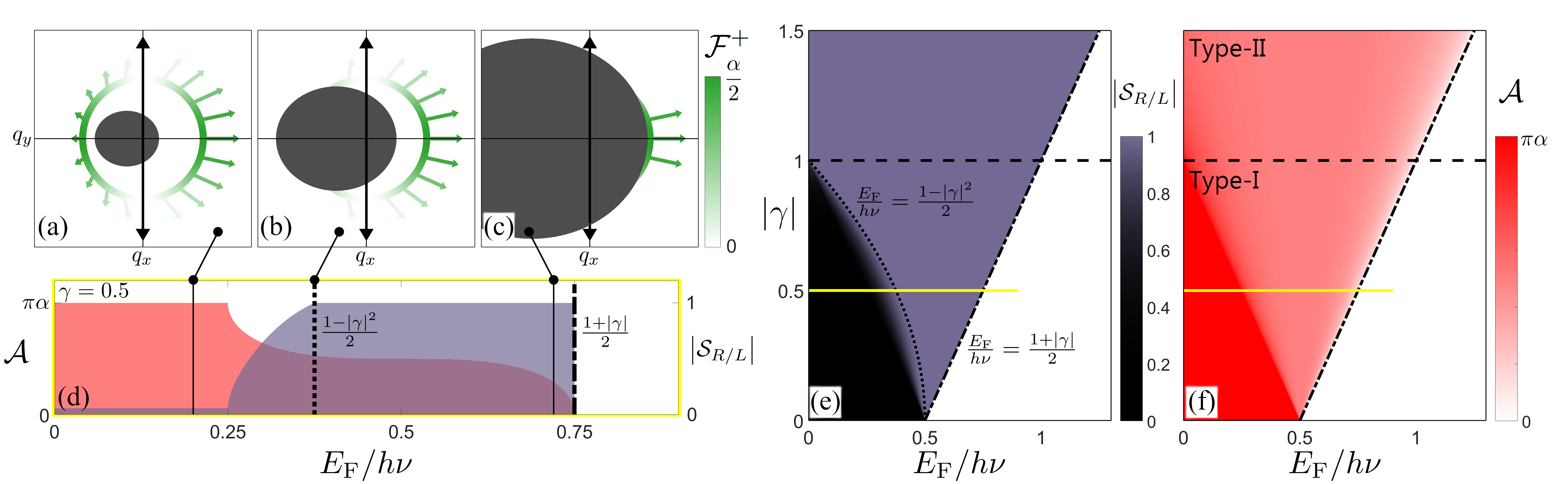}
    \caption{(a)-(c) Distribution of photoexcited carriers ($\mathcal{F}^+$) in a single valley ($\xi = +$) of a type-I Dirac cone material for photons polarized along the crystallographic $\hat{\textbf{y}}$ axis for a range of Fermi energies. Regions of Pauli blocked transitions are shaded in gray. In this plot the tilt parameter is $\gamma = 0.5$, the anisotropy parameter is $\eta = 1$ and $\alpha \approx 1/137$ is the fine structure constant. The group velocities of photoexcited carriers have been projected in to wavevector space and sketched as arrows. In panels (b) and (c), carriers with valley index $\xi = +$ ($-$) will propagate along the $\textbf{x}$ ($-\hat{\textbf{x}}$) direction - displaying spatial separation of valley carriers. On the contrary, in panel (a) the carriers move in each direction irrespective of their valley index. (d) The total absorption $\mathcal{A}$ (red) and the absolute value of the valley polarization degree $\left| \mathcal{S}_{R/L} \right|$ (purple) are plotted as functions of the Fermi level ($E_\text{F}$) normalized by photon energy ($h \nu$). In panels (e) and (f), the degree of valley polarization and the absorption are plotted as functions of $E_\text{F}/h \nu$ and $\left|\gamma\right|$. The contours $E_\text{F} = h \nu(1 - \left|\gamma\right|^2)/2$ (dotted) and $E_\text{F} = h \nu (1 + \left|\gamma\right|)/2$ (dot-dashed) correspond to the lines in panel (d).}
    \label{fig:SPlot}
\end{figure*}
\subsection{Type-I Dirac cones}
Initially, we consider a type-I ($\left| \gamma \right| < 1$) Dirac cone material with Fermi level sitting above the Dirac point (see Fig.\,\ref{fig:PolarisationChange}). For incident photons polarized along the crystallographic $\hat{\textbf{y}}$ axis ($\theta = \pi/2$), the transition rate dictates that photoexcited electrons are created close to the tilt ($q_x$) axis of the Dirac cone. However, any state inside the Pauli blocked regions does not undergo a transition. The resulting distribution of photoexcited carriers shows that if the tilt parameter takes on a positive value ($\gamma > 0$) the majority of carriers in valley $\xi = +$ ($-$) are created on the right (left) side of the Dirac cone [see Figs.\,\ref{fig:PolarisationChange}(a) and (b)]. The group velocities resulting from the conical band structure dictate that at photocreation, carriers with valley number $\xi = +$ will propagate to the right ($\hat{\textbf{x}}$ direction) whilst carriers with valley number $\xi = -$ will propagate to the left ($-\hat{\textbf{x}}$ direction) towards the different sides of the illuminated light spot. We note that in general, the tilt parameter could take a negative value ($\gamma < 0$), in this case carriers from valley $\xi$ will propagate in the $-\xi\hat{\textbf{x}}$ direction. 

Although we have highlighted spatial separation of valley carriers for a specific polarization ($\theta = \pi/2$), this phenomenon occurs, to a lesser extent, for all polarizations. As the polarization plane is rotated towards the crystallographic $\hat{\textbf{x}}$ axis ($\theta = 0$) an increased amount of carriers move along the $\hat{\textbf{y}}$ axis, nevertheless, there is still a significant amount of valley separation [see Figs.\,\ref{fig:PolarisationChange}(c) and (d)]. As it is possible to determine the orientation of the crystallographic axes of a Dirac semimetal with an optical procedure\,\cite{Wild2022}, aligning the incident photon polarization close to $\hat{\textbf{y}}$ will yield optimal results.

By tuning a gate-voltage we can modify the Fermi level which via Pauli blocking changes the distribution of photoexcited carriers. If the Fermi level sits close to the Dirac point then little to no transitions are Pauli blocked and carriers from both valleys propagate in all directions, minimizing spatial separation of valley carriers [see Fig.\,\ref{fig:SPlot}(a)]. As the gate-voltage is increased, the regions of Pauli blocked transitions grow and we obtain considerable spatial separation of valley carriers [see Fig.\,\ref{fig:SPlot}(b)]. If we were to increase the gate-voltage further, our absorption would decrease [see Fig.\,\ref{fig:SPlot}(c)] before stopping altogether. 

To quantify the valley separation we define the parameter $\mathcal{N}_{R(L)}^\xi$ to be the percentage of photoexcited carriers in valley $\xi$ that propagate to the right (left) side of the light spot along the crystallographic $\hat{\textbf{x}}$ axis
\begin{equation}
   \mathcal{N}^\xi_R = \frac{\int_\Sigma \mathcal{F}^\xi(\varphi_\textbf{q})\text{d} \varphi_\textbf{q}}{\int \mathcal{F}^\xi(\varphi_\textbf{q})\text{d} \varphi_\textbf{q}}.
\end{equation}
The domain of integration $\Sigma$ is defined as the set of angles $\varphi_\textbf{q}$ corresponding to a positive $\hat{\textbf{x}}$ component of the group velocity $v^\xi_{x} = (1/\hbar)\partial_{q_x}E^\xi_+$ where $\mathcal{F}^\xi$ is defined by Eq.\,(\ref{eq:Carrier}). The parameter $\mathcal{N}^\xi_L$ can be deduced from the normalization condition $\mathcal{N}_R^\xi + \mathcal{N}_L^\xi = 1$. Using these quantities, we can define the degree of valley polarization at the right-hand side of the light spot as
\begin{equation}
\label{eq:SFunction}
    \mathcal{S}_R = \frac{\mathcal{N}^+_{R} - \mathcal{N}^-_{R}}{\mathcal{N}^+_{R} + \mathcal{N}^-_{R}}.
\end{equation}
We provide an analytic expression for the valley polarization degree at either side of the light spot ($\mathcal{S}_{R/L}$) in Appendix\,\ref{sec:SFormula}. If all photoexcited carriers at the right-hand side of the light spot are from valley $\xi$ then the valley polarization takes on the value $\mathcal{S}_R = \xi$, in contrast, if there is an equal number of carriers from either valley then $\mathcal{S}_R = 0$. The valley polarization at the left-hand side of the light spot is the opposite of the right-hand side $\mathcal{S}_L = -\mathcal{S}_R$. The degree of valley polarization can be detected when photoexcited carriers propagate into a nearby gapped material, where they can recombine emitting circularly polarized photons with handedness determined by their valley index $\xi$. The degree of valley polarization maps on to the degree of circular polarization of the emitted light.

\begin{figure}[t]
    \centering
    \includegraphics[width=0.49\textwidth]{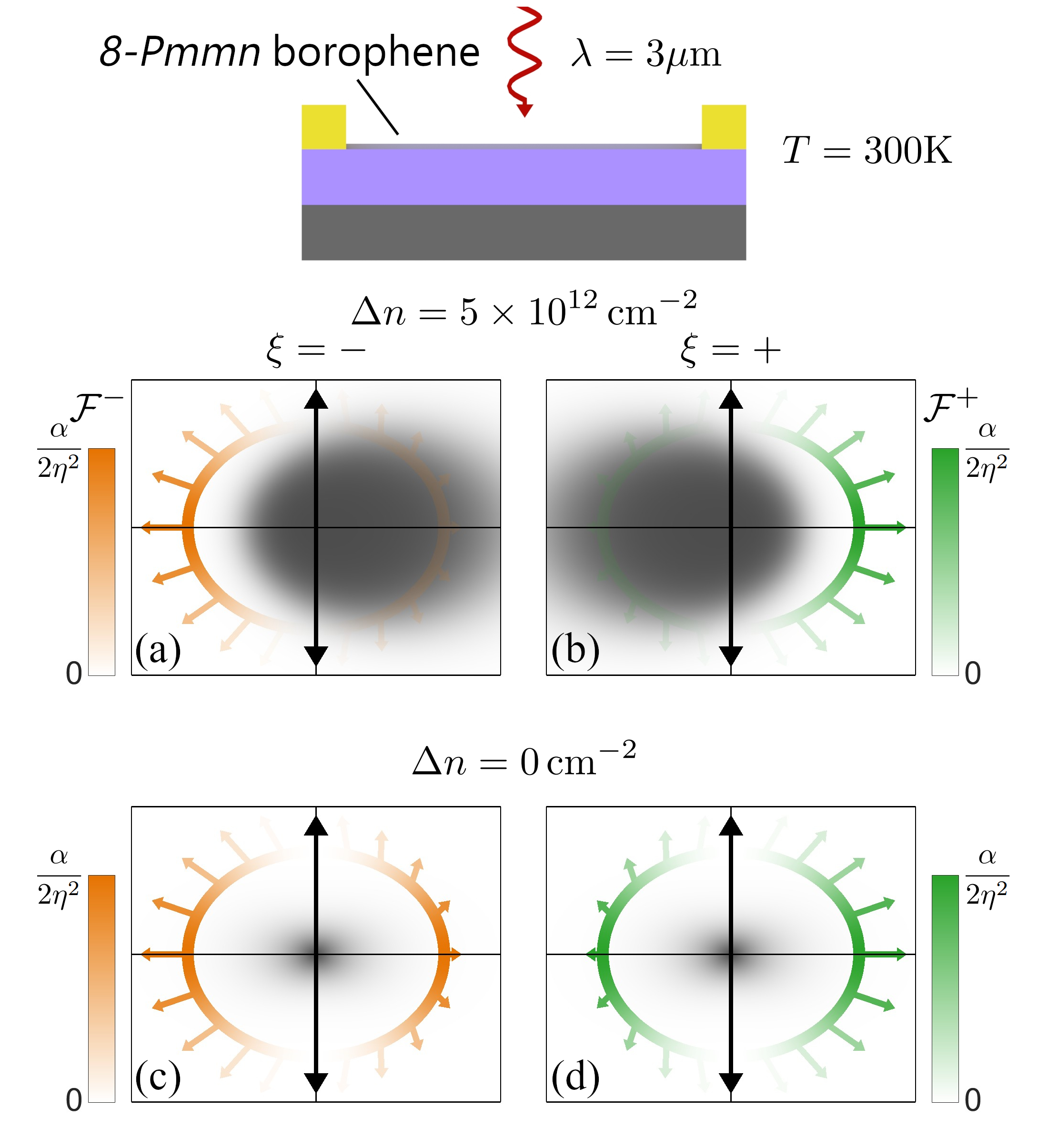}
    \caption{Distribution of photoexcited carriers ($\mathcal{F}^\xi$) in the candidate type-I Dirac cone material $8-Pmmn$ borophene. In this plot the Dirac cone tilt is $\gamma = 0.46$, the anisotropy parameter is $\eta = 0.80$ and $\alpha \approx 1/137$ is the fine structure constant. The monolayer is incident upon by infrared photons polarized along the crystallographic $\hat{\textbf{y}}$ axis with wavelength $\lambda = 3\mu\text{m}$ at ambient temperature $T = 300\text{K}$. The finite temperature blurs the regions of Pauli blocked transitions, in conjunction with previous figures, the stronger the Pauli blocking the more opaque the gray shading. By utilizing a back-gate configuration, carriers can be added to the monolayer moving the Fermi level. The carrier density $\Delta n$ is defined as the density of carriers added from charge neutrality ($\Delta n = 0$). The group velocities of photoexcited carriers have been projected in to wavevector space and sketched as arrows.}
    \label{fig:FermiEnergyChange}
\end{figure}We can now quantify the degree of valley polarization for a specific tilted Dirac cone geometry with different Fermi energies [see Fig.\,\ref{fig:SPlot}(d)]. In type-I Dirac cones the degree of valley polarization is maximal ($\left| \mathcal{S}_{R/L} \right| = 1$) for gate-voltages greater than or equal to $h \nu (1 - \left| \gamma \right|^2)/2$ [see Fig.\,\ref{fig:SPlot}(e)]. However, we note that for there to be any absorption, the Fermi energy must also be less than $h \nu (1 + \left| \gamma \right|)/2$ [see Fig.\,\ref{fig:SPlot}(f)]. Therefore, it is theoretically possible to achieve perfect valley carrier separation ($\left| \mathcal{S}_{L/R} \right| = 1$ and $\mathcal{A} > 0$) in type-I Dirac cones for gate voltages $h \nu (1 - \left| \gamma \right|^2)/2 \leq E_\text{F} < h \nu (1 + \left| \gamma \right|)/2$ as demonstrated for a specific value of tilt in Figs.\,\ref{fig:SPlot}(b) and (c). This bound clearly vanishes in the limit of $\gamma = 0$ which emphasizes that this mechanism of spatial separation of valley carriers is not possible in non-tilted Dirac cone materials such as graphene.

\begin{figure}[t]
    \centering
    \includegraphics[width=0.48\textwidth]{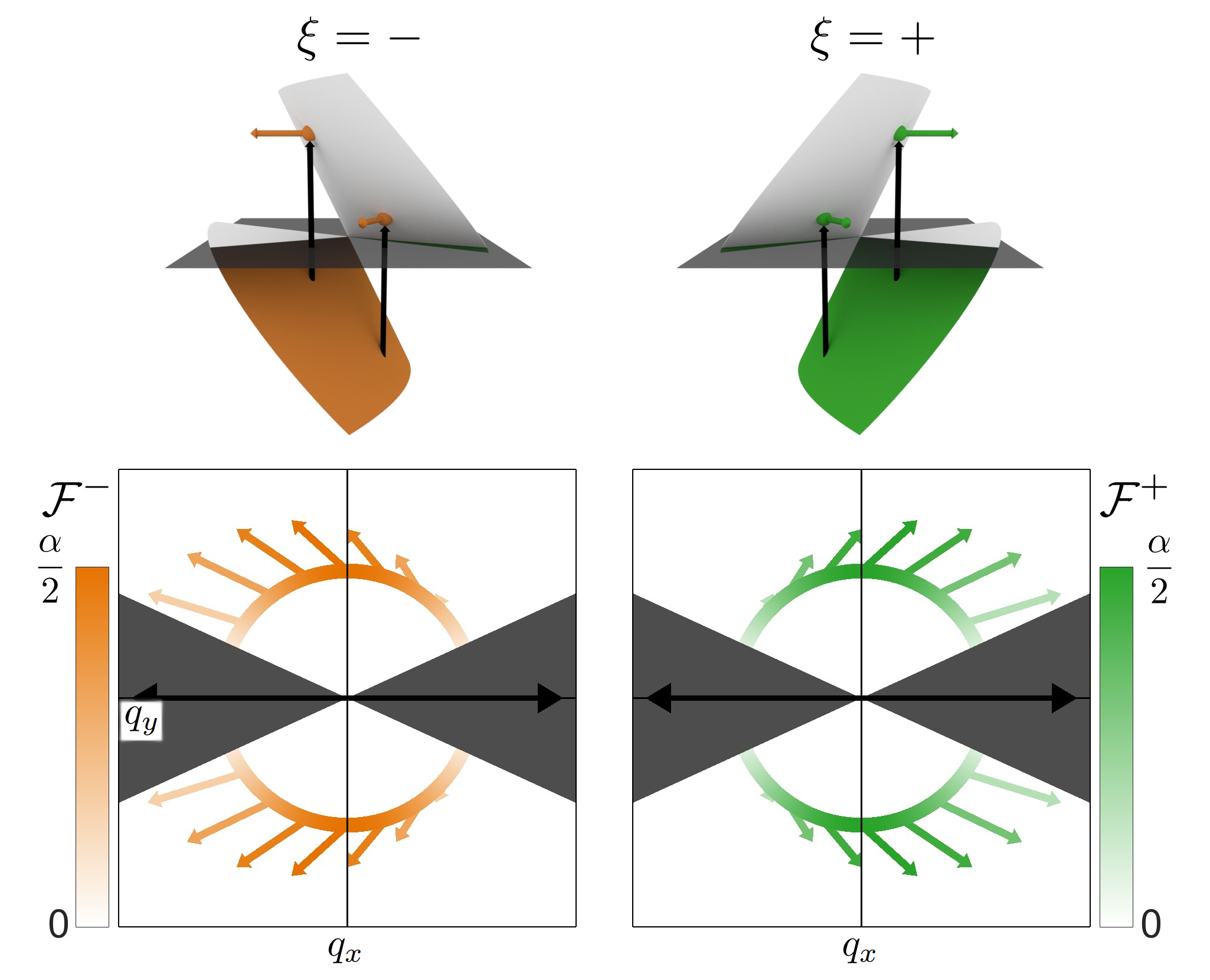}
    \caption{Distribution of photoexcited carriers ($\mathcal{F}^\xi$) in a type-II Dirac cone material with valley indices $\xi = +$ ($-$) sketched in green (orange). In this plot the Dirac cone tilt is $\gamma = 1.1$, the anisotropy parameter is $\eta = 1$ and $\alpha \approx 1/137$ is the fine structure constant. Regions of Pauli blocked transitions are shaded in gray. The group velocities of photoexcited carriers have been projected in to wavevector space and sketched as arrows. Due to the super-critically tilted band structure, all photoexcited carriers are spatially separated according to their valley index.}
    \label{fig:typeII}
\end{figure}
\subsection{Special case: $8-Pmmn$ borophene}
As a specific case study of our theory we demonstrate the spatial separation of valley carriers in the predicted tilted type-I Dirac cone material $8-Pmmn$ borophene under illumination of infrared photons. In this material the Dirac cones have a Hamiltonian of the form given in Eq.\,(\ref{eq:Hamiltonian}) with Fermi velocity $v_\text{F} = 8.6 \times 10^5\text{ms}^{-1}$, tilt parameter $\gamma = 0.46$ and anisotropy parameter $\eta = 0.80$\,\cite{PhysRevB.94.165403}. It can be seen that with $8-Pmmn$ borophene at room temperature ($T = 300\text{K}$) it will be possible to achieve valley separation by adding carriers to the system via a back-gate configuration [see Figs.\,\ref{fig:FermiEnergyChange}(a) and (b)]. By removing carriers, bringing the Fermi level back to charge neutrality, valley separation can be turned off [see Figs.\,\ref{fig:FermiEnergyChange}(c) and (d)].

\subsection{Type-II Dirac cones}
Unlike their type-I counterparts, type-II ($\left| \gamma \right| > 1$) Dirac cones are super-critically tilted. The group velocity of these Dirac cones dictates that all photoexcited carriers will be spatially separated according to their valley index (see Fig.\,\ref{fig:typeII}). In other words, as long as there is absorption [which requires $E_\text{F} < h \nu (1 + \left| \gamma \right|)/2$] there will always be full spatial separation of valley carriers ($\left| \mathcal{S}_{R/L}\right| = 1$) for any polarization of light [see Figs.\,\ref{fig:SPlot}(e) and (f)]. As we always have full spatial separation of valley carriers in type-II Dirac cones, for demonstrative purposes, we pick the polarization of light that maximizes the absorption and number of carriers which corresponds to $\theta = 0$.

\subsection{Carrier relaxation enhanced momentum alignment in type-III Dirac cones}
Up until this point, critically tilted type-III Dirac cones have merely marked the boundary between type-I and II Dirac cones. However, when including the effects of carrier relaxation, type-III Dirac cones offer an interesting mechanism of momentum alignment not possible in any other tilted Dirac cones. 

\begin{figure}[t]
    \centering
    \includegraphics[width=0.49\textwidth]{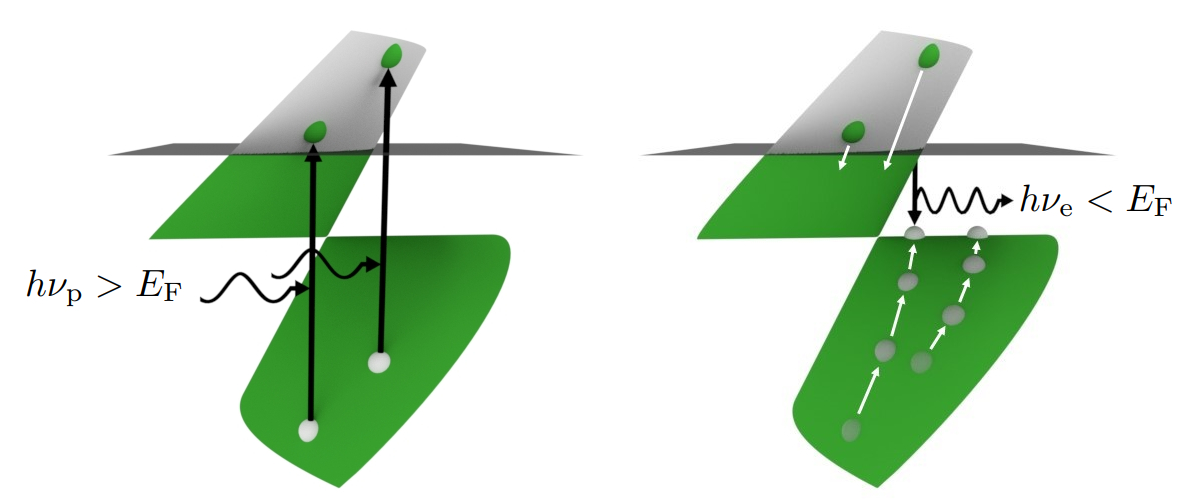}
    \caption{Schematic of enhanced momentum alignment in a type-III Dirac cone with valley index $\xi = +$ and tilt parameter $\gamma = 1$. Black arrows indicate interband absorption/emission and white arrows indicate relaxation via carrier-carrier and carrier-phonon scattering processes. After interband absorption $h \nu > E_\text{F}$ holes float towards the Fermi level becoming trapped in an intermediate state with wavevector $q_y = 0$. Upon recombination photons will be emitted with polarization aligned with the crystallographic $\hat{\textbf{y}}$ axis and energy $h \nu < E_\text{F}$.}
    \label{fig:TypeIII}
\end{figure}

Critically tilted type-III ($\left| \gamma \right| = 1$) Dirac cones have a peculiar band structure in which the extrema of the upper and lower bands are one-dimensional lines in the wavevector space. First, we pump the material with arbitrarily polarized photons with energy $h \nu_\text{p} > E_\text{F}$ (see Fig.\,\ref{fig:TypeIII}). The resulting electrons and holes relax via a combination of carrier-carrier and carrier-phonon scattering processes to the most energetically favorable state. The holes aim to increase their energy, floating to the one-dimensional band maxima. The holes become stranded in these intermediate states which are perfectly aligned in momenta. Any holes that relaxed to a small wavevector $\mid \! q_x \! \mid$ will be able to recombine with electrons in the upper band emitting photons of energy $h \nu_\text{e} < E_\text{F}$. Due to the momentum alignment of these holes, the emitted photons will have polarization aligned with the crystallographic $\hat{\textbf{y}}$ axis. This mechanism of emission via an intermediate state is known as hot luminescence\,\cite{REBANE1978223}. By modifying the Fermi level with a back-gate voltage, the emission energy of these photons can be tuned to the terahertz regime yielding a highly-polarized tunable terahertz emitter.

\section{Conclusion}
The realization of the valley-polarized currents via the valley Hall effect provided the elementary building block for valleytronic devices in gapped Dirac cone materials (see review articles\,\cite{C4NR01600A,Xu2914,Schaibley2016,Mak2016,C7CS00210F,RevModPhys.90.021001} and references therein). This discovery sparked a desire for valleytronic components that in conjunction with the valley Hall effect could lead to valley-sensitive logic gates for classical and quantum computing applications\,\cite{https://doi.org/10.1002/smll.201801483}. In our work we demonstrate the spatial separation of valley carriers away from the light spot in gapless Dirac materials with tilted Dirac cones. Our discovery paves the way to the realization of novel valleytronic devices benefiting from the superior transport properties of massless Dirac fermions.

With the recent burst of interest in massless tilted Dirac cone materials there have been several theoretical works investigating the valley-dependent transport of carriers traversing gated junctions, waveguides and external fields\,\cite{PhysRevB.97.235440,PhysRevB.97.235113,Islam_2018,Zheng_2020,ng2021mapping}. Combining these transport techniques with the optical spatial separation of valley carriers proposed in our work could enable the design of valleytronic components such as valley filters and switches in gapless materials. It may also be possible to further direct the propagation of valley carriers across graphene-based interconnects based on electrostatic waveguides \,\cite{PhysRevB.81.245431,ZALIPAEV2013216}, quantum wire leads \,\cite{PhysRevB.76.045433} or gated junctions in externally applied fields\,\cite{PhysRevB.99.205431}. Furthermore, the spatial separation of valley carriers in gapless tilted Dirac cone materials could be combined with valley-sensitive components of gapped Dirac cone materials such as valley transistors\,\cite{PhysRevB.86.165411} or decoding the valley index via emission of circularly polarized light\,\cite{PhysRevLett.99.236809,PhysRevB.77.235406,PhysRevB.103.165415}. This would require a detailed understanding of the transport phenomena occurring at the interface between gapless and gapped Dirac cone materials. It is well-known that placing graphene on a hexagonal boron nitride substrate induces a superlattice structure inducing local regions with pseudo-gaps\,\cite{2014NatPh..10..451W,10.1038/ncomms7308, PhysRevB.103.115406} - a similar technique for tilted Dirac materials should enable the seamless transport of valley carriers between gapless and gapped regions in the spectrum allowing valley index measurement. Lastly, the theoretical and computational predictions of two-dimensional materials hosting massless tilted Dirac cones are rapidly growing in number\,\cite{PhysRevLett.112.085502,PhysRevB.93.241405,PhysRevB.94.165403,doi:10.1143/JPSJ.75.054705,PhysRevB.78.045415,PhysRevLett.105.037203,doi:10.1126/science.1256815,PhysRevX.6.041069,PhysRevB.94.195423,Ma2016,https://doi.org/10.1002/pssr.201800081,PhysRevB.98.121102,PhysRevB.100.235401,PhysRevB.100.205102,PhysRevB.102.041109}. The experimental efforts aiming at realizing these materials are catching up \,\cite{doi:10.1126/science.aad1080,Feng2016,doi:10.1098/rsos.181605}. We hope that the prospect of optovalleytronics put forward in our work will stimulate further research into massless tilted Dirac cone materials.
\section*{ACKNOWLEDGMENTS}
This work was supported by the EU H2020-MSCA-RISE projects TERASSE (Project No. 823878) and DiSeTCom (Project No. 823728) as well as by the NATO Science for Peace and Security project NATO.SPS.MYP.G5860. A.W. is supported by a UK EPSRC PhD studentship (Ref. 2239575). E.M. acknowledges financial support from the Royal Society
(Grant No. IEC/R2/192166).

\appendix

\section{Analytic expression for the distribution of photoexcited carriers in tilted Dirac cones} 
\label{sec:FFormula}
In this Appendix, we present the expression for the distribution of photoexcited carriers $\mathcal{F}^\xi(\varphi_\textbf{q})$ for carriers with valley index $\xi$ as a function of wavevector angle $\varphi_\textbf{q}$. Combining Eqs.(\ref{eq:eigenvalues}) and (\ref{eq:DeltaE})-(\ref{eq:Carrier}) and solving the resultant integral yields
\begin{widetext}
\begin{equation}
\label{eq:CarrierGeneralTAniso}
\begin{split}
    \mathcal{F}^{\xi}(\varphi_\textbf{q}) =& \frac{\alpha}{2}\frac{\eta^2 \sin^2(\varphi_\textbf{q} - \theta)}{\big[ \eta^2 \cos^2(\varphi_\textbf{q}) + \sin^2(\varphi_\textbf{q}) \big]^2} \Bigg( 1 - \Bigg\{ 1 + \text{exp} \Bigg[  \frac{h \nu \xi \eta \gamma\cos(\varphi_\textbf{q})}{2 k_\text{B} T \sqrt{\eta^2 \cos^2(\varphi_\textbf{q}) + \sin^2(\varphi_\textbf{q})}} + \frac{h \nu}{2 k_\text{B} T} - \frac{\mu}{k_\text{B} T} \Bigg] \Bigg\}^{-1}\Bigg)\times\\
   & \Bigg\{ 1 + \text{exp} \Bigg[  \frac{h \nu \xi \eta \gamma  \cos(\varphi_\textbf{q})}{2 k_\text{B} T \sqrt{\eta^2 \cos^2(\varphi_\textbf{q}) + \sin^2(\varphi_\textbf{q})}} - \frac{h \nu}{2 k_\text{B} T} - \frac{\mu}{k_\text{B} T} \Bigg] \Bigg\}^{-1}.
\end{split}
\end{equation}
\end{widetext}

\section{Analytic expression for the polarization of valley carriers}
\label{sec:SFormula}
In this Appendix, we provide an analytic expression for the valley polarization of photoexcited carriers that propagate to the right-hand side of the light spot. As the Dirac cones in either valley are tilted in opposite directions, the percentage of carriers propagating to the right in valley $\xi$ is equal to the percentage of carriers propagating to the left in valley $-\xi$ yielding the identity $\mathcal{N}^\xi_R = \mathcal{N}^{-\xi}_L$. Utilizing this expression, and the identity $\mathcal{N}^\xi_R + \mathcal{N}^\xi_L = 1$, Eq.\,(\ref{eq:SFunction}) can be simplified to $\mathcal{S}_R = \xi(\mathcal{N}_R^\xi - \mathcal{N}_L^\xi)$ which can be defined through the distribution of photoexcited carriers as
\begin{equation}
\label{eq:ValleyPolarisation}
    \mathcal{S}_R = \xi \frac{\int_0^{2\pi} \mathcal{F}^\xi(\varphi_\textbf{q}) \text{sign}\big[ v^\xi_{x}(\varphi_\textbf{q})\big] \text{d} \varphi_\textbf{q}}{\int_0^{2\pi} \mathcal{F}^\xi(\varphi_\textbf{q}) \text{d} \varphi_\textbf{q}},
\end{equation}
where $\text{sign}(...)$ is the sign function and $v^\xi_{x}(\varphi_\textbf{q})$ is the $\hat{\textbf{x}}$ component of the group velocity. The valley polarization at the left-hand side of the light spot is related to the right-hand side by the expression $\mathcal{S}_L = -\mathcal{S}_R$. Solving Eq.\,(\ref{eq:ValleyPolarisation}) for type-I ($\left|\gamma\right| < 1$) Dirac cone materials incident upon by photons polarized along the crystallographic $\hat{\textbf{y}}$ axis yields the degree of valley polarization $\mathcal{S}_R = \text{sign}(\gamma)\mathcal{S}$ where
\begin{equation}
    \mathcal{S} = \begin{cases}
    \mathcal{S}_0, & \frac{E_\text{F}}{h \nu} \leq (1 - \left|\gamma\right|)/2,\\
    \mathcal{S}_1, & (1 - \left|\gamma\right|)/2 < \frac{E_\text{F}}{h \nu} < (1 - \left|\gamma\right|^2)/2,\\
   1, & (1 - \left|\gamma\right|^2)/2 \leq \frac{E_\text{F}}{h \nu} < (1 + \left|\gamma\right|)/2,
    \end{cases}
\end{equation}
with
\begin{equation}
    \mathcal{S}_0 = \frac{2}{\pi} \Big[ \arccos(-\left|\gamma\right|) - \left|\gamma\right| \sqrt{1 - \left|\gamma\right|^2}\Big] - 1,
\end{equation}
and 
\begin{equation}
    \mathcal{S}_1 = 2 \Bigg[\frac{\arccos(-\left|\gamma\right|) - \left|\gamma\right| \sqrt{1 - \left|\gamma\right|^2}}{\arccos(\psi) + \psi\sqrt{1 - \psi^2}}\Bigg] - 1,
\end{equation}
where $\psi = (1/\left|\gamma\right|)(2E_\text{F}/h \nu - 1)$. In this expression we have assumed the low temperature limit where $k_\text{B} T$ is small compared to the photon and Fermi energies.

For type-II ($\left|\gamma\right| > 1$) Dirac cones, assuming there is absorption which requires $E_\text{F} < h \nu (1 + \left|\gamma\right|)/2$, the degree of valley polarization is equal to unity ($\left| \mathcal{S}_{R/L} \right| = 1$).

\bibliography{apssamp}

\end{document}